\documentstyle[twocolumn,aps]{revtex}
\begin{document}
\title{
Scaling Properties of Conductance at Integer Quantum Hall Plateau
Transitions
}
\author{
Xiaosha Wang, Qiming Li, and  C. M. Soukoulis
}
\address{
Ames Laboratory-USDOE and Department of Physics and Astronomy \\
Iowa State University, Ames, Iowa 50011\\
}
\author{\parbox[t]{5.5in}{\small
We investigate the scaling properties of zero temperature
conductances at integer quantum Hall plateau transitions
in the lowest Landau band of a two-dimensional tight-binding model.
Scaling is obeyed for all energy and system sizes with critical exponent 
$\nu \approx \frac {7}{3}$. The arithmetic average of the conductance at the
localization-delocalization
critical point is found to be $<G>_c = 0.506 \frac {e^2}{h}$, in agreement
with the universal longitudinal conductance $<\sigma_{xx}> = \frac {1} {2}
\frac {e^2}{h}$ 
predicted by an analytical theory. 
The probability distribution of the conductance at the critical point is 
broad with a dip at small G. 
 }}

\maketitle

The transitions between the integer quantum Hall (IQH) plateaus are
believed to be a manifestation of the localization to delocalization
transition in two dimensional electron systems in the presence of
a strong magnetic field. The phenomenon\cite{review} is characterized
at finite temperatures by the appearance of a conductance peak as the 
Hall conductance varies continuously between the
precisely quantized values with the change of applied magnetic 
field\cite{review}.
The existence of both extended and localized states
is required to describe this fascinating phenomenon.
Following extensive experimental and theoretical investigation, 
a consistent picture has emerged on the nature of this remarkable transition.
In a two dimensional disordered non-interacting
electron system,  extended states do not exist as a result of Anderson
localization
except at a singular energy near the center of each of the Landau
sub-bands\cite{chalker,huck}.
The localization length diverges at these critical energies
as $\xi \sim |E-E_c|^{-\nu}$, signifying a continuous zero-temperature
quantum phase
transition. Electron conduction close to the transition
is controlled by the ratio of the coherent length over the localization
length\cite{pru}. As the system size increases or, equivalently, as
the temperature decreases to zero for macroscopic samples, the system
scales to stable Hall insulator fixed points characterized by
$(\sigma_{xx}, \sigma_{xy}) =( 0, n)\frac{e^2}{h}$.

Although a consensus has reached on the general picture, several
fundamental issues remain unsolved. One such issue, the
universality of the transition,  has attracted much
attention recently\cite{univ}. Based on Chern-Simons formalism,
Lee-Zhang-Kivelson\cite{klz}
predicted a transition between quantized plateaus
$(\sigma_{xx}, \sigma_{xy}) =( 0, n)\frac{e^2}{h}$
and $(0, n+1)\frac{e^2}{h}$ with universal critical points $(\frac {1}{2},
n+\frac {1}{2})\frac{e^2}{h}$, independent of the microscopic details of any
models. Such universality at the quantum critical point has also
been suggested previously\cite{univ}. The universality of the critical
exponents appears to have been born out in all the latest numerical
studies with different microscopic models\cite{chalker,huck,bhatt,gammel}.
These works produced values consistent
with experimental measurement\cite{wei1,koch,wei} and close to the
analytical predicted value\cite{mil} $\nu = \frac {7}{3}$
for quantum percolation. The universality of $\sigma_{xx}$, however, is
still controversial. Extrapolation of dynamical conductivity from Kubo
formalisms\cite{gammel} and from spectra function calculations\cite{bhatt}
on short ranged potentials produced values in reasonable agreement with
$\sigma^c_{xx} = \frac {1} {2}$ at the critical point.  A recent extensive and 
direct numerical study\cite{wang}on mesoscopic systems with the network 
model\cite{chalker}, however, produced  $G_c =(0.58 \pm 0.03) \frac {e^2} {h}$, 
in disagreement with the analytical prediction and previous numerical
results. Given the
fact that the two terminal conductance G is not exactly the same as the
longitudinal 
conductivity $\sigma_{xx}$, the small but significant difference appears to
be resolved\cite{wang} 
in favor of a new universal value for the two terminal conductance $<G_c>$.
Experimental 
attempts\cite{exptgc} to measure the conductance directly have not produced
consistent values.

Unlike the Hall conductance which is exactly quantized as a consequence of 
topological invariance, the longitudinal conductance itself is a
sample-dependent quantity for
mesocopic systems with electron coherent lengths exceeding the
sample dimension.  The universality of the critical conductance in principle 
applies only to macroscopic systems in which self-averaging is expected.
It is not clear at the present what average procedure will produce
the experimentally observed quantity. Fluctuation has been the hallmark
of quantum transport in mesoscopic systems\cite{lee} and its characteristics
provide
important clues on the system as a whole.  It is therefore equally
interesting to investigate 
the probability distribution of the conductance. In particular, it is
important to ask whether, under appropriate conditions, the distribution
of the conductance P(G) approaches a limiting universal form.
In three-dimensional disorder-driven metal-insulator transitions, strong
evidence exist on the universality of the conductance distribution in the
metallic and insulating phases as well as at the critical
point\cite{shapiro,slevin}.
The physical arguments for the existence of such a universality can be equally 
applied to the localization-delocalization transition in two dimensional
systems\cite{wang,raikh}. Conductance distributions at
IQH plateau transitions have been investigated for
the network model only\cite{cho}. The results compared well with the recent
experimental measurement on mesoscopic systems\cite{cobden}.

In this paper, we investigate the scaling properties of the conductance as well
as the critical conductance and its distribution in a two dimensional
system described by a tight-binding model. This is perhaps the simplest
system that exhibits the correct critical behavior of the
localization-delocalization transition. To our knowledge, no such calculations
have been previously reported for the tight-binding model. Our aim is to
determine
accurately the critical zero temperature
two-terminal conductance in the tight-binding model and to see whether there
is universality, at least between the tight-binding model and the
network model investigated by Wang et al.\cite{wang}. Our results are in
agreement with 
the scaling hypothesis and a critical exponent of $\nu = \frac {7}{3}$. More
importantly, based on finite
size scaling analysis, we obtain $<G_c>=0.506$ at the critical point,
very close to the value expected from analytical theory $G_c = \frac {1} {2}$,
but in disagreement with the recent extensive numerical study on the network 
model\cite{wang}. This difference, if persists, has important implications
on the universality of the two terminal critical conductance.   
We have also analyzed the
distribution of the conductance at the critical point. The probability distribution of the two terminal conductance be
The moments of the distribution are
surprisingly similar to that of the network model.

The tight-binding Hamiltonian is given as follows
\begin{equation}
H =\sum_i \epsilon_i |i><i| + \sum_{<ij>}(t_{ij}|i><j| + c.c.)
\end{equation}
where $\epsilon$ is a  random site energy uniformly distributed within
[-W/2, W/2].
The complex hopping integral $t_{ij}$ carries the phase due to applied
magnetic fields via the standard Peierls substitution,
\begin{equation}
t_{ij}=t_0e^{{-\frac{i2\pi e} {hc}}\int_i^j {\vec{A}d\vec{l}}}.
\end{equation}
The sum is carried over the nearest neighbor sites $<ij>$  only. $t_0$
is taken as the unit of energy. Periodic boundary conditions are applied
in the transverse direction. In continuum,  quantized Landau levels under 
a magnetic field are broadened into Landau bands by impurity scattering.
In lattice models, Landau bands form even in ordered systems as a result of
degeneracy
breaking.  These bands will be further broadened by disorder. The lowest
Landau band in 
the tight-binding model has been shown\cite{review} to describe the quantum Hall
transition well.

To calculate the zero-temperature two terminal conductance numerically, 
we employ the transfer matrix method which obtains the final transmission
matrix by multiplications and inversions of transfer matrix. 
The disordered square sample of size MxM is 
sandwiched between two perfect leads of the same width. Both the sample and the 
leads are governed by the tight-binding Hamiltonian (1). No disorder exists
in the leads.
The two terminal conductance is then given by the following multichannel
Landuaer 
formula\cite{g}
\begin{equation}
G = \frac{e^2} {h}Tr(T^\dagger T)
\end{equation}
where T is the total transmission matrix through the disordered sample
with the propagating channels in the leads as basis.  Keep in mind that G
defined 
here is for one spin only.

For the purpose of investigating the scaling and the critical conductance,
we have chosen a fixed magnetic field such that the flux per square is one
eighth
of the flux quantum (f =1/8) and a disorder strength W=4. If universality
persists, 
both the critical exponent $\nu$ and the critical conductance $<G_c>$ are
expected 
not to depend on the applied field and the disorder strength. Based on a
previous
finite size scaling study on the localization length, the critical point of
the lowest 
Landau level at this field and disorder strength is accurately known to be
at E=-3.40.
At this disorder, all Landau bands show substantial band-mixing
except 
the lowest one. For disordered mesoscopic systems at zero
temperature, electrons propagate through the entire sample without being
scattered inelastically by phonons.  Scattering by random static
impurity, on the other hand, produces configuration dependent conductance
fluctuations. We present in Figure 1(a) the averaged conductance 
$<G>$ in the lowest 
Landau band  for different system sizes. The conductance clearly peaks at
the critical 
energy $E_c$=-3.40 and falls rapidly away from $E_c$. As the system size
increases, the conductance curve become narrower and the peak conductance 
increases. The 
continuing narrowing of the width of the conductance curve
as system size increases indicates that in the macroscopic limit, only 
the states at the critical energy can transport electrons across the 
sample, in agreement with the conventional picture that all states 
are localized except at the center of the band\cite{review}.

An important property is the scaling of the conductance G as a function
of the system size. According to the finite size scaling idea,
the conductance is expected to be determined solely by the ratio of
the localization length to the system dimension M close to critical point.
However, there is known irrelevant finite size corrections such that the
scaling is modified as
\begin{equation}
G(E,M) = G(E_c,M) f(\xi(E)/M)
\end{equation}
where $\xi(E)$ is the macroscopic localization length at energy E and
f(x) is a universal function.  The size dependence of the conductance 
maximum $G(E_c,M)$ represents the irrelevant finite size corrections,
\begin{equation}
G_{s}(E_c,M)=G_c - a M^{-y_{irr}}.
\end{equation}
Utilizing $\xi  \sim |E-E_c|^{-\nu}$, we obtain the expression
\begin{eqnarray}
G(E,M) = G(E_c,M) f(|E-E_c|^{-\nu}/M)\nonumber \\
=G(E_c,M) F(|E-E_c| M^{1/\nu}).
\end{eqnarray}
Should scaling exists in our system as expected, then all of our data
for different E and M would collapse on one curve provided the correct
values of $E_c$ and $\nu$ are chosen. The results of such a scaling
procedure are shown in Figure 1(b) for
arithmetic average with E$_c$=-3.40 and the best fit $\nu$=2.37. 
Scaling behavior is clearly established.
Deviation for M=16 and for higher energy E is due to the
finite size effect and the effect of mixing with higher bands, respectively.
Thus scaling of conductance around the critical points indicates
a critical exponent consistent with a universal value $\nu =\frac {7}{3}$. 
The same scaling is obeyed for the geometric average, shown in Fig. 1(c) 
with exactly the same critical exponent.

Of central importance is the exact value of the conductance at the critical
point. As mentioned before, due to the finite size correction from irrelevant
fields, the conductance at the critical point depends on system size.
In Figure 2, we present the arithmetic and geometric averaged conductance
at the critical point, $E_c=-3.40$.
$G(E_c,M)$ increases with increasing size M and will eventually saturate
at the critical conductance $G_c$ for macroscopic systems.
This is in contrast to the constant amplitude 
ratio\cite{review} at $E_c$ for the finite size localization length.
We mention that in order to achieve good statistics, more
than 10,000 samples are taken for M up to 160 and 8000 samples for M=192.
To extrapolate to macroscopic systems, we have fitted our data to
Eq. (5) with a least square fit (shown as lines in Figure 3). The most
likely fit is determined
by minimizing the $\chi^2$ statistics\cite{recipe}
\begin{equation}
\chi^2 = \sum_i  (\frac {G(E_c,M_i)-G_{s}(E_c,M_i)} {\sigma_i})^2,
\end{equation}
where the summation i is over all the system sizes and $\sigma_i$ is 
the standard deviation of 
$<G(E_c,M)>$.  We obtain $G_c =0.506$ and $y_{irr}=0.72$ as the best fit with
a goodness of fit
Q=0.12. Fits with Q larger than 0.1 are believable. We have also used the
projection method\cite{recipe}
to estimate the confidence limit.  Brackets of $[0.499, 0.511]$ and $[0.495,
0.517]$ are 
obtained with a confidence level of 95.4$\%$ and 99.73$\%$, respectively.
Our results agree 
with the analytical assertion that the universal longitudinal conductance
$G_c$ is $\frac {1} {2}$. 
The difference between our results
for the tight-binding
model and the equally extensive results for the network model\cite{wang},
if persists, could indicate that the critical two-terminal 
conductance is not universal.
The extrapolated value for the 
geometric average is $<G_c>_g=0.438$ with 
$y_{irr}=0.52$ and  
a goodness of fit Q=0.50. An interesting remark is that from
the reported value of higher moments of G, 
we infer that in the network model, the geometrically
averaged conductance  $<G>_g \approx 0.5$.

The distribution of the conductance also shows interesting properties.
At the critical point. this distribution is broad and ranges between
0 and 1. Fluctuations, as measured by the standard deviation, is of the
same order of magnitude as the average conductance itself. For localized
state, the distribution is known to be Poisson-like. At 
critical points, it has been proposed that the conductance 
distribution should be universal
independent of the size of the system. This assertion is based on the fact
that there is no length scale since the localization length
diverges at  the critical point. However, we know there is
non-negligible finite size corrections
to the scaling, as shown in the analysis of the $G_c$. Thus the
conductance distribution at the critical point do show size
dependences (Figure 3a-d). The probability distribution is broad, ranging
from 0 to 1. It also shows progressive development of a dip around G=0 
as the whole distribution flattens. 
For two dimensional systems, analytical
descriptions of the statistical distribution of the conductance is lacking.
Eventually for very large sizes, the distribution
saturate to the final, presumably universal, distribution for
macroscopic systems. Calculation of higher moment of the conductance,
$<G^n>$, results in values 0.08, 0.013, 0.003 and 0.001 for n=2, 4, 6 and 8, 
respectively.  These values closely resemble the results reported by 
Wang et al.\cite{wang}.  The probability distribution at the 2D quantum 
critical point is quite different from that of the 3D
systems\cite{shapiro,slevin}.
We also point out that the distribution in lnG, although not gaussian, has
much better central tendency. The universality of these distributions 
will be examined in future work.

Ames Laboratory is operated for the U. S. Department of Energy by Iowa
State University under contract No. W-7405-ENG-82. This work was supported
by the Director of Energy Research, Office of Basic Energy Science, and
Advanced Energy Projects.


\begin{figure}
\caption{
Average conductance $<G>$ in the lowest landau Band. (a) Conductance vs
energy for M=16, 32, 64, 96, 128, 160, and 192.  Normalized conductance
as a function of scaled variable $x= |E-E_c|M^{1/\nu}$ with $E_c =- 3.40$ and
$\nu = 2.37$ for the arithmetic (b) and the geometric (c) average. The number
of samples
for each data point ranges from 50 for M=192 and 10000 for M=16. 
}
\end{figure}

\begin{figure}
\caption{
Conductance at the critical point, $<G(E_c,M)>$,  as a function of
system size M for square samples of $M \times M$. The lines are least square
fits to $G_s(E_c, M)=G_c-aM^{-y_{irr}}$.
The error bars are smaller than the size of the symbols.
}
\end{figure}

\begin{figure}
\caption{
Distribution of the conductance G at the critical point $E_c=-3.40$ for
different sample sizes. a) M=16, b) M=32, c) M=64, and d) M=128. Each size
has more than 10000 samples. Distributions
at M=192  (not shown here) is almost identical with that of M=128 within the
statistical 
fluctuation. 
}
\end{figure}


\begin{references}
\bibitem{review}
For reviews, see B. Hukestein, Rev. Mod. Phys. {\bf 67}, 357 (1995); {\it
The Quantum Hall Effect},
edited by R.~E.~Prange and S.~M.~Girvin (Springer-Verlag, New York, 1990).
\bibitem{chalker}
J.~T. Chalker and P.~D.~Coddington, J. Phys. C{\bf 21}, 2665 (1988).
\bibitem{huck}
B.~ Hukesterin and B.~Kramer, Phys. Rev. Lett. {\bf 64}, 1437 (1990).
\bibitem{pru}
A.~M.~M. Pruisken, Phys. Rev. Lett. {\bf 61}, 1297 (1988).
\bibitem{univ}
X.-G.~Wen and A.~Zee, Int. J. Mod. Phys. B{\bf 4}, 437 (1990);
M.~P.~A.~Fisher, G.Grinsterin, and S.~M.~Girvin, Phys. Rev. Lett. {\bf 64},
587 (1990).
\bibitem{klz}
S.~A.~Kivelson, D.-H.~Lee, and S.-C.~Zhang, Phys. Rev. B{\bf 46}, 2223 (1992).
\bibitem{bhatt}
Y.~Huo, E.~Hetzel, and R.~N.~Bhatt, Phys. Rev. Lett. {\bf 70}, 481 (1993).
\bibitem{gammel}
B.~M.~Gammel and W.~Brenig, Phys. Rev. Lett. {\bf 73}, 3286 (1994).
\bibitem{wei1}
H.~P.~Wei, D.~C.~Tsui, M~.Paalanen, and A.~M.~M. Pruisken, Phys. Rev. Lett.
{\bf 61}, 1294 (1988).
\bibitem{koch}
S.~Koch, R.~J.~Huag, K.~von Klitzing, and K.~Ploog, Phys. Rev. Lett. {\bf
67}, 883 (1991).
\bibitem{wei}
H.~P.~Wei, L.~W.~Engel, and D.~C. Tsui, Phys. Rev. B{\bf 50}, 14609 (1991).
\bibitem{mil}
G.~V.~Mil'nikov and I.~M.~Sokolov, Pis'ma Zh. Eksp. Teor. Fiz. {\bf 48}, 494
(1988) [JETP Lett., {\bf 48}, 536 (1988)].
\bibitem{wang}
Z.~Wang, B.~Jovanovic, and D.-H.~Lee, Phys. Rev. Lett. {\bf 77}, 4426 1996).
\bibitem{exptgc}
D. Shahar et al.  Phys. Rev. Lett. {\bf 74}, 4511 (1995); and L.~P.
Rokhinson et al.,
Sol. Stat. Comm., {\bf 96}, 309 (1995).
\bibitem{lee}
For review, see P. A. Lee and T. V. Ramakrishnan, {\it Rev. Mod.
Phys.} {\bf 57}, 287 (1985).

\bibitem{shapiro}
A. Cohen and B. Shapiro, Int. J. of Mod. Phys. {\bf 6}, 1243 (1992).
\bibitem{slevin}
K. Slevin and T. Ohtsuki, Phys. Rev. Lett. {\bf 78}, 4083 (1997).
\bibitem{raikh}
A.~G.~Galstyan and M.~E.~Raikh, cond=mat/9701010 (1997).
\bibitem{cho}
S.~Cho and M.~P.~A.~Fisher, Phys. Rev. B {\bf 55 }, 1637 (1997).
\bibitem{cobden}
D.~H.~Cobden and E.~Kogan, Phys. Rev. B{\bf 54}, R17316 (1996).
\bibitem {g}
J. L. Pichard and G. Andre, Europhys. Lett. {\bf 2} 477 (1986);
D. S. Fisher and P. A. Lee, {\it Phys. Rev. B} {\bf 23} 685 (1981);
and E. N. Economou and C. M. Soukoulis, {\it Phys. Rev. Lett.} {\bf
46}, 618 (1981)
\bibitem{shen}
D.~N. Shen and Z.~Y. Weng, Phys. Rev. Lett.{\bf 78}, 318 (1997).
\bibitem{recipe}
{\it Numerical Recipes in Fortran}, edited by W. Press, B. Flannery, and S.
Teukolsky (Cambridge
University Press, Cambridge, England, 1992), Chap. 15. 
\end{references}
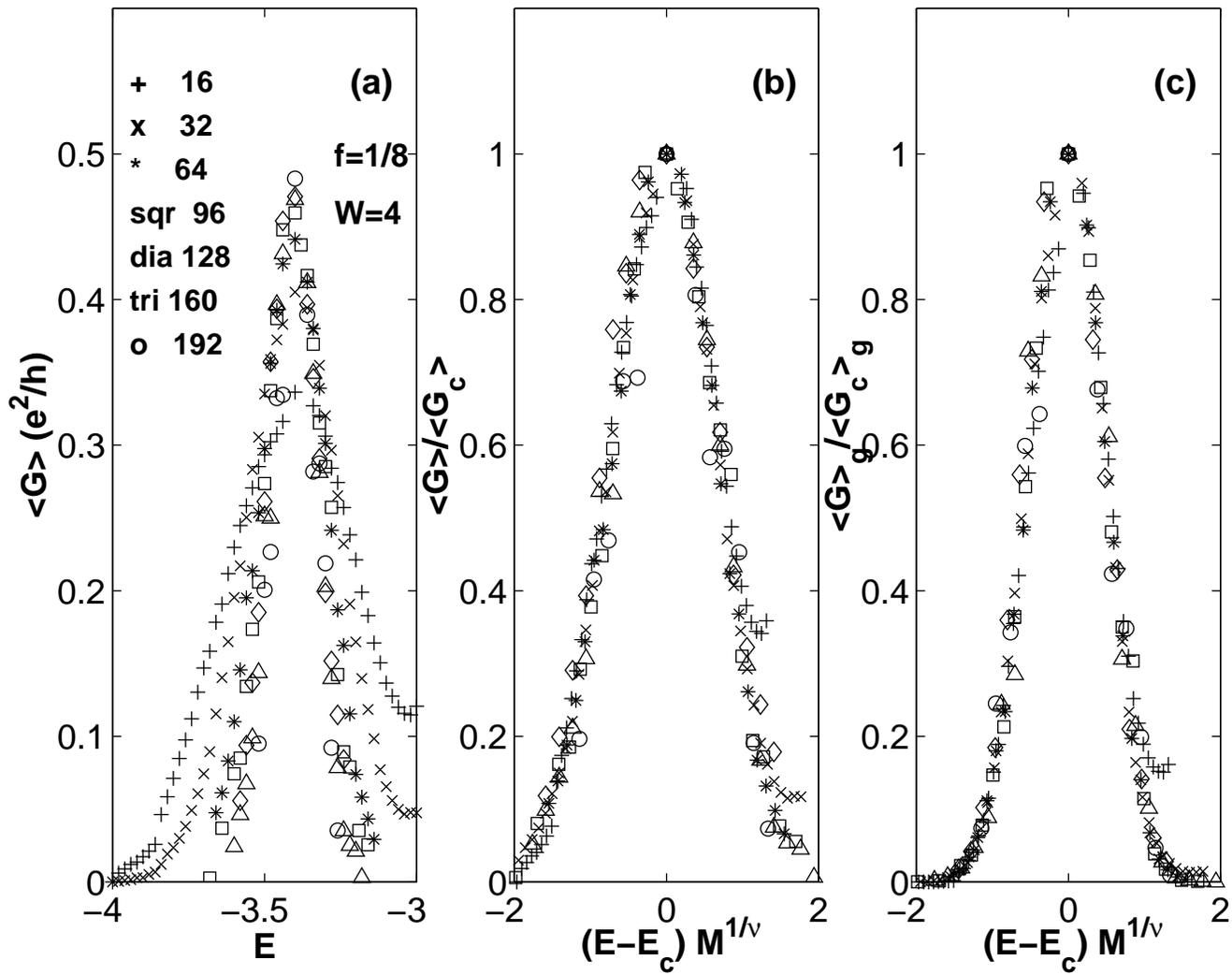
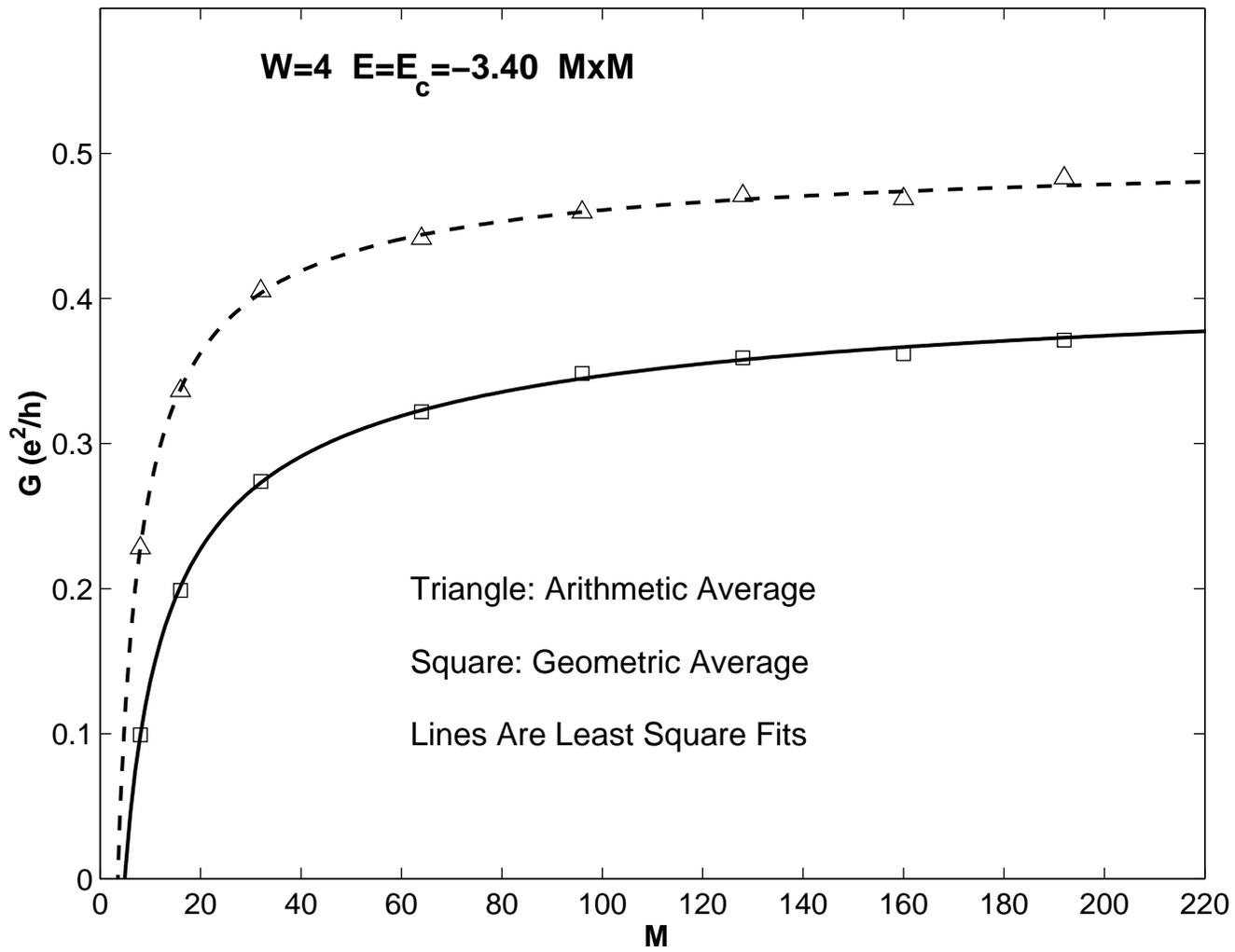
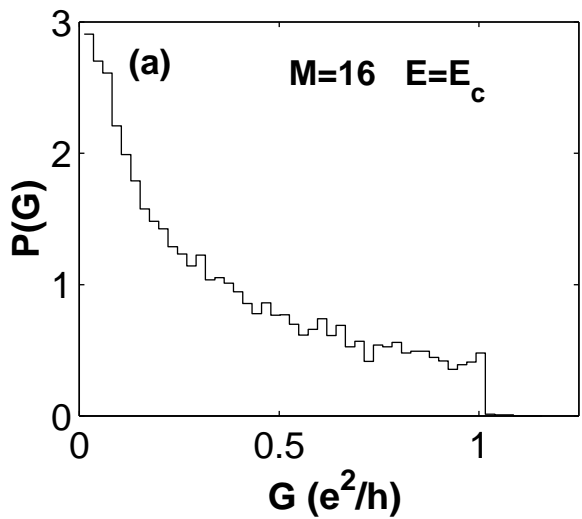
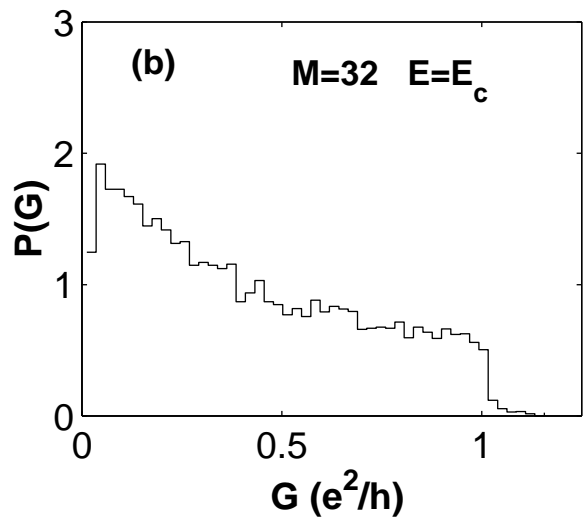
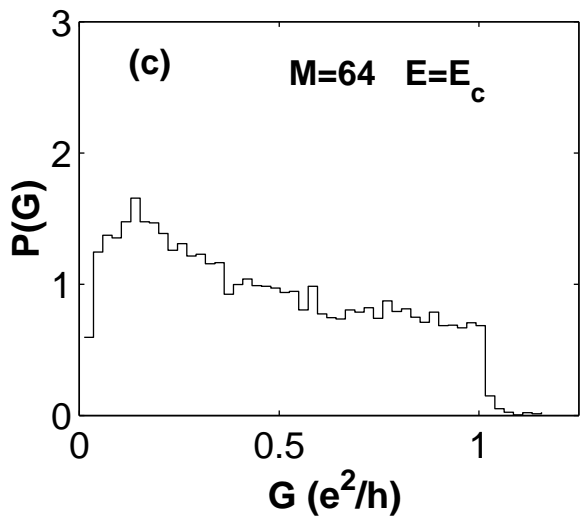
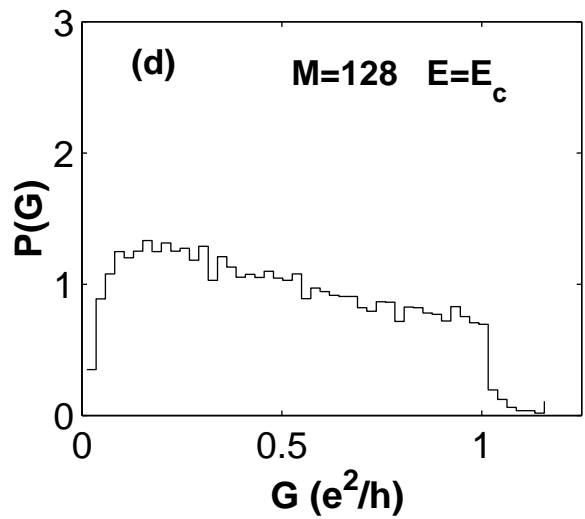
\end{document}